\documentclass[10pt,prl,nofootinbib,notitlepage,twocolumn,superscriptaddress]{revtex4-1}
\usepackage{amsmath}
\usepackage{amssymb,amsfonts}
\usepackage{graphicx}
\graphicspath{{pdfpics/}}
\usepackage{mathtools}
\usepackage{relsize}
\usepackage{lmodern}
\usepackage{slantsc}
\usepackage{scalefnt}
\usepackage{subfigure}
\usepackage{dsfont}
\usepackage{xspace}
\usepackage{boxedminipage}
\usepackage{ifpdf}
\usepackage{pgffor}
\usepackage{xifthen}
\usepackage{tensor}
\usepackage{array}
\usepackage{multirow}
\usepackage{tikz}
\usetikzlibrary{shapes.geometric}
\usepackage{color}
\usetikzlibrary{arrows,matrix,calc,scopes,decorations.markings}
\allowdisplaybreaks[1]
\usepackage{times}
\usepackage{hyperref}
\usepackage{epstopdf}
\usepackage{enumerate}

\usepackage{color}
\definecolor{dred}{rgb}{.8,0.2,.2}
\definecolor{ddred}{rgb}{.8,0.5,.5}
\definecolor{dblue}{rgb}{.2,0.2,.8}
\definecolor{dgreen}{rgb}{.2,0.5,.2}

\newcommand{\be}{\begin{equation}}
\newcommand{\ee}{\end{equation}}
\newcommand{\bse}{\begin{subequations}}
\newcommand{\ese}{\end{subequations}}
\newcommand{\ket}[1]{|{#1}\rangle}

\newcommand{\bbra}[1]{\left\langle{#1}\right|}
\newcommand{\bracket}[2]{\langle{#1}\vert{#2}\rangle}
\newcommand{\Z}{\mathbb{Z}}

\newcommand{\bpm}{\begin{pmatrix}}
\newcommand{\epm}{\end{pmatrix}}
\newcommand{\bmm}{\begin{matrix}}
\newcommand{\emm}{\end{matrix}}

\newcommand{\x}{\times}

\allowdisplaybreaks

\makeatletter
\newcommand*{\Relbarfill@}{\arrowfill@\Relbar\Relbar\Relbar}
\newcommand*{\xeq}[2][]{\ext@arrow 0055\Relbarfill@{#1}{#2}}
\makeatother

\newcommand{\Psix}[3][1]{
\begin{tikzpicture}[scale=0.8]
\node[name=s, regular polygon, regular polygon sides=6, minimum size=1cm, outer sep=0pt ,draw] at (0,0) {}; 
\foreach \anchor/\x/\y /\xx/\yy /\b in
{corner 1/0.17/0.17*1.732/-0.11/0.18/1, corner 2/-0.17/0.17*1.732/0.07/0.18/2, corner 3/-0.34/0/-0.15/-0.18/3, corner 4/-0.17/-0.17*1.732/-0.22/-0.05/4, corner 5/0.17/-0.17*1.732/0.2/-0.05/5, corner 6/0.34/0/0.15/-0.18/6}
{
 \draw[shift=(s.\anchor)] (0,0) -- (\x,\y) node at(\xx,\yy) {$#2_{\text{\scalebox{0.7}{$\b$}}}$};
 \ifnum #1=1
 \draw[shift=(s.\anchor),<-,>=stealth', line width=0.01pt] (s.\anchor) -- (\x,\y);
 \fi
 }
\foreach \anchor/\xx/\yy /\a in
{side 1/0/-0.18/1, side 2/-0.18/0.05/2, side 3/0.15/0.05/3, side 4/0/-0.18/4, side 5/-0.18/0.05/5, side 6/0.15/0.05/6}
 \draw[shift=(s.\anchor)]  node at(\xx,\yy) {$#3_{\text{\scalebox{0.7}{$\a$}}}$};
\ifnum #1=1{
  \foreach \anchorr/\anchorf in
   {corner 1/corner 2, corner 2/corner 3, corner 3/corner 4, corner 4/corner 5, corner 5/corner 6, corner 6/corner 1}
   \draw[shift=(s.\anchorr), ->, >=stealth', line width=0.01pt]  (s.\anchorr) -- (s.\anchorf);}
 \else {
  \foreach \anchorb/\anchorw in
   {corner 1/corner 2, corner 3/corner 4, corner 5/corner 6} {
   \node[fill=black, circle, minimum size=0, inner sep=0, outer sep=0, draw] at(s.\anchorb) {};
   \node[fill=white, circle, minimum size=0, inner sep=0, outer sep=0, draw] at(s.\anchorw) {};}
}
\fi
\end{tikzpicture}
}

\begin{document}

\title{Experimental Identification of Non-Abelian Topological Orders on a Quantum Simulator}

\author{Keren Li}
\affiliation{State Key Laboratory of Low-Dimensional Quantum Physics and Department of Physics, Tsinghua University, Beijing 100084, China}
\affiliation{Institute for Quantum Computing and Department of Physics and Astronomy,
University of Waterloo, Waterloo N2L 3G1, Ontario, Canada}

\author{Yidun Wan}
\affiliation{Department of Physics and Center for Field Theory and Particle Physics, Fudan University, Shanghai 200433, China}
\affiliation{Collaborative Innovation Center of Advanced Microstructures, Nanjing University, Nanjing 210093, China}
\affiliation{Perimeter Institute for Theoretical Physics, Waterloo N2L 2Y5, Ontario, Canada}

\author{Ling-Yan Hung}
\affiliation{State Key Laboratory of Surface Physics and Department of Physics, Fudan University, 220 Handan Road, Shanghai 200433, China}
\affiliation{Department of Physics and Center for Field Theory and Particle Physics, Fudan University, Shanghai 200433, China}
\affiliation{Collaborative Innovation Center of Advanced Microstructures, Nanjing University, Nanjing 210093, China}

\author{Tian Lan}
\affiliation{Perimeter Institute for Theoretical Physics, Waterloo N2L 2Y5, Ontario, Canada}

\author{Guilu Long}
\affiliation{State Key Laboratory of Low-Dimensional Quantum Physics and Department of Physics, Tsinghua University, Beijing 100084, China}

\author{Dawei Lu}
\email{d29lu@uwaterloo.ca}
\affiliation{Institute for Quantum Computing and Department of Physics and Astronomy,
University of Waterloo, Waterloo N2L 3G1, Ontario, Canada}

\author{Bei Zeng}
\affiliation{Institute for Quantum Computing and Department of Physics and Astronomy,
University of Waterloo, Waterloo N2L 3G1, Ontario, Canada}
\affiliation{Department of Mathematics \& Statistics, University of
  Guelph, Guelph N1G 2W1, Ontario, Canada}
\affiliation{Canadian Institute for Advanced Research, Toronto M5G 1Z8,
  Ontario, Canada}

\author{Raymond Laflamme}
\affiliation{Institute for Quantum Computing and Department of Physics and Astronomy,
University of Waterloo, Waterloo N2L 3G1, Ontario, Canada}
\affiliation{Perimeter Institute for Theoretical Physics, Waterloo N2L 2Y5, Ontario,
Canada}
\affiliation{Canadian Institute for Advanced Research, Toronto M5G 1Z8,
  Ontario, Canada}%

\begin{abstract}
Topological orders can be used as media for topological quantum computing --- a promising quantum computation model due to its invulnerability against local errors. Conversely, a quantum simulator, often regarded as a quantum computing device for special purposes, also offers a way of characterizing topological orders. Here, we show how to identify distinct topological orders via measuring their modular $S$ and $T$ matrices. In particular, we employ a nuclear magnetic resonance quantum simulator to study the properties of three topologically ordered matter phases described by the string-net model with two string types, including the $\Z_2$ toric code, doubled semion, and doubled Fibonacci. The third one, non-Abelian Fibonacci order is notably expected to be the simplest candidate for universal topological quantum computing. Our experiment serves as the basic module, built on which one can simulate braiding of non-Abelian anyons and ultimately topological quantum computation via the braiding, and thus provides a new approach of investigating topological orders using quantum computers.
\end{abstract}
\maketitle

\textit{Introduction ---} Beyond the Landau-Ginzburg paradigm of symmetry breaking, topologically orders describe gapped quantum phases of  matter  with a myriad of properties depending only on the topology but not of any microscopic details of the host system \cite{Landau1950,Wen1989a,Wen1990a,Wen1990,Chen2012a}. These properties are thus robust against local perturbations. Two such properties are a finite set of degenerate ground states and a corresponding set of gapped (non-Abelian) anyon excitations \cite{Kitaev2003a,Kitaev2006}.  While the former may lead to a robust quantum memory \cite{Dennis2002}, the latter may form a logical space that supports quantum computation via the unitary braiding of the anyons \cite{Kitaev2003a,Freedman2003,Stern2006,Nayak2008}. This architecture of quantum computation is called topological quantum computation (TQC), because the ground states, anyons, and braiding operations are nonlocal by nature and hence are invulnerable against local errors. The most promising and simplest candidate topological order for universal TQC is the Fibonacci order \cite{Stern2006,Nayak2008}, which bears a non-Abelian anyon species $\tau$, and the braiding operations of two or more $\tau$'s form a universal set of unitary gates.

The potential, paramount applications of topological orders urges studies of topological orders in real systems. Rather than directly realizing a topological order in a real system, simulating it on a quantum computer offers an alternative means of investigating topological orders, where the first step is naturally to identify distinct topological orders. A topological order has three key features: topology-protected ground state degeneracy (GSD), finite number of anyon types, and topological properties of the anyons\cite{Levin2004}. Particularly, the third characteristic, topological properties of the anyons, includes the self-statistics, braiding, and fusion of the anyons. The self-statistics of an anyon can be a fraction, recorded by the modular $T$ matrix of a topological order, which in a proper basis is diagonal. Meanwhile, the braiding of two anyons can be captured by an observable called $S$-matrix. The fusion of two anyons is an interaction that produces other (not necessarily different) anyons in the topological order, which is also captured by the $S$ matrix. Therefore, two distinct topological orders with the same GSD can still be distinguished by comparing their modular $T$ and $S$ matrices \cite{Kitaev2006,Bonderson2006,Bonderson2006a,Bonderson2007,Rowell2009}.

In this work, we consider the string-net model, also known as the Levin-Wen model\cite{Levin2004}, with only two string types. In this case, the model describes only three topological orders. The first two are the $\Z_2$ toric code and doubled semion order, which are Abelian topological orders. The third is the doubled Fibonacci order, which is non-Abelian and the candidate for universal TQC. As these three topological orders possess the same GSD on a torus, we need to identify them via their modular matrices. In experiment, we simulate each of the three topological orders on a nuclear magnetic resonance (NMR) quantum simulator \cite{somaroo1999quantum,tseng1999quantum,peng2005quantum,du2010nmr,alvarez2010nmr,lu2011simulation}, and measure its modular transformation $ST^{-1}$ as a whole. As each of the three topological orders possess a unique $ST^{-1}$, we have thus identified all three topological orders of the string-net model in practice, and our experiment opens up a new way of identifying topological orders using quantum simulators.

\begin{figure}[!ht]
\centering
\includegraphics[scale=0.1]{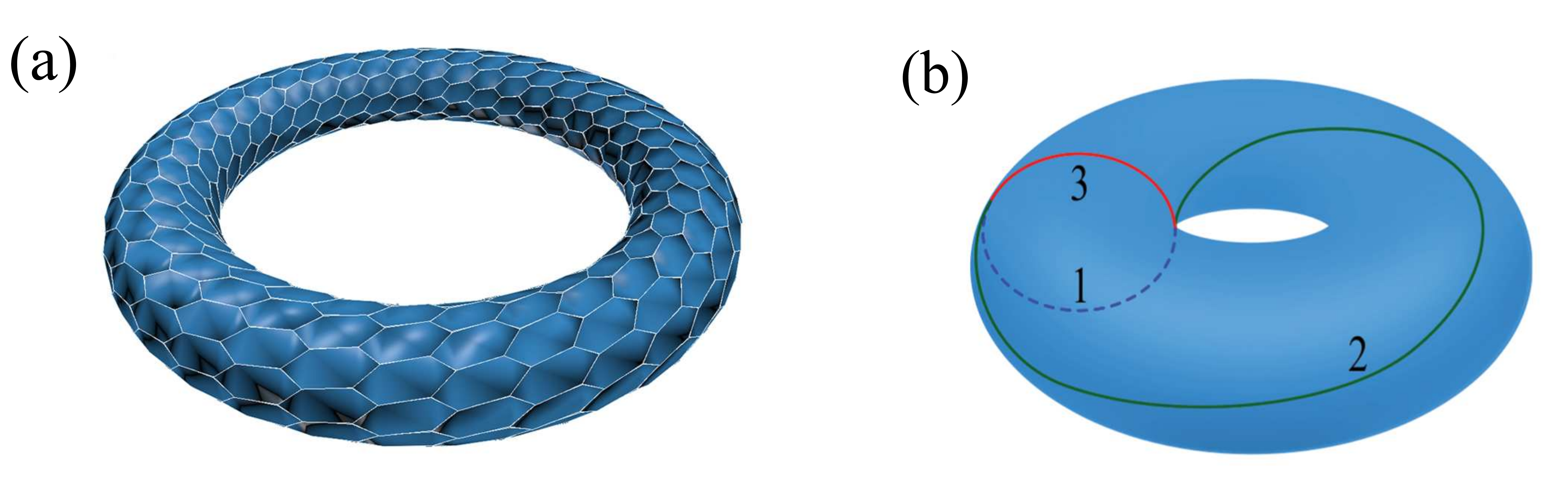}
\caption{(a) Honeycomb lattice on a torus. For the sake of ground states only, this can always be simplified into (b) The minimal honeycomb lattice on a torus: three edges (strings) labeled by $1$, $2$, and $3$, two trivalent vertices, and one plaquette --- the entire torus.}
\label{fig:minTorus}
\end{figure}

\textit{String-net model ---}
String-net models are exactly solvable, infrared fixed point, effective models of topological orders in two spatial dimensions \cite{Levin2004}. A string-net model is specified by a set of input data: string types $\{i,j,k,\dots\}$, fusion rules $\{N^k_{ij}\in\Z_{\geq 0}\}$, and a Hamiltonian $H$, all defined on the honeycomb lattice (Fig. \ref{fig:minTorus}(a)). The strings are the fundamental degrees of freedom of the model, and each edge of the lattice has a unique string type, which can evolve under the Hamiltonian. For example, strings may be thought as spins living on the edges of the lattice. A fusion rule $N^k_{ij}$ is defined on a vertex where the three incident edges of the string types $i,j$, and $k$, via the equation $i\x j=\sum_{k}N^k_{ij}k$, in which the non-negative integers $N^k_{ij}$ are fusion coefficients. In this work, we deal with the cases where the strings are self-dual, i.e., $i=i^*$ for all string type $i$'s, and $N^k_{ij}\in\{0,1\}$ only. The Hamiltonian of this model reads
\be\label{eq:snH}
H=-\sum_v A_v-\sum_p B_p,
\ee
where the sums are respectively over all the vertices and plaquettes of the honeycomb lattice. It turns out that all the $A_v$ and $B_p$ operators commute with each other, which renders the model exactly solvable. More importantly, all these operators are projectors and thus have eigenvalues either zero or one. We direct readers to the supplemental material for a detailed description of the string-net model \cite{supple}.

Given a set of input data of the string-net model, the degenerate ground states, anyon excitations, and modular $T$ and $S$ matrices form the set of output data, which characterizes a specific topological order based on the input data.

\textit{Minimal honeycomb lattice on a torus ---}
On a torus, as far as ground states are concerned, one can always shrink the lattice to the minimal honeycomb lattice with merely three edges, two vertices, and one plaquette as shown in Fig. \ref{fig:minTorus} by the so called $F$-moves \cite{Levin2004,Hung2012,Hu2012,supple}.
The total Hilbert space is spanned by the basis states $\ket{123}$, where the numbers label both the edges and the string types carried respectively on the edges. These basis states are orthonormal: $\bracket{1'2'3'}{123}=\delta_{1,1'}\delta_{2,2'}\delta_{3,3'}$. The next step is to find the matrix form of the string-net Hamiltonian in Eq. \eqref{eq:snH} on this minimal honeycomb lattice. As we only concern about the ground states, we can set $A_v=1$ at both vertices of the minimal lattice. The only nontrivial part of the Hamiltonian is thus $B_p$ on the sole plaquette. We can derive that the matrix elements of $B_p$ are \cite{supple}
\begin{align}
& \bbra{1''2''3''}B_p\ket{123}\label{eq:Bp}\\
=& \frac{1}{D}\sum_{s,1',2',3'}d_s F^{123}_{s3'2'}F^{231}_{s1'3'}F^{3'12'}_{s2''1'} F^{3'1'2}_{s2'1''}F^{2''3'1'}_{s1''3''}F^{1''2'3'}_{s3''2''}.\nonumber
\end{align}
Here, the strings $1'$, $2'$, and $3'$ being summed over are those in the intermediate states, and $s$ represents an average over all possible string types associated with the action of $B_p$. $F^{ijm}_{kln}$ are the $F$-symbols, which for the case with $\kappa$ string types, are a collection of  $\kappa^6$  complex numbers determined by the fusion rules. The quantity $d_i=F^{ii0}_{ii0}$, where $0\leq i \leq \kappa-1$, is defined as the quantum dimension (not the actual dimension of any Hilbert space but a convenient notation) of the string type $i$. Since the quantum dimensions $d_i$ are defined by the $F$-symbol normalization, one may instead specify the quantum dimensions as part of a set of input data. In our setting, all $F$-symbols are real and $\kappa=2$.

When $\kappa=2$ where a string can precisely be simulated by a qubit, $B_p$ is an $8\x 8$ real matrix, and so is the entire Hamiltonian.
Meanwhile, there are three and only three possible sets of fusion rules, each of which gives rise to a string-net Hamiltonian describing a distinct topological order \cite{supple}. In the following, we only list the defining facts and topological properties of the three topological orders for $\kappa=2$ but leave certain details such as the matrix forms of the Hamiltonian to the supplemental material \cite{supple}. For each topological order, the types of anyons, basis of the ground states, and $T$ and $S$ matrices are shown in Table. \ref{topoorder}.

1. $\Z_2$ \textit{toric code}. The input data includes two string types $0$ and $1$, fusion rules $0\x 1=1$ and $1\x 1=0$, and quantum dimensions $d_0=d_1=1$. The ground state space is four-dimensional on the torus.
The four types of anyons are $1$, $e$, $m$, and $\epsilon$, where $e$ and $m$ are self-bosons but mutual fermions, and $\epsilon$ is a fermion. The set of output data characterizes the $\Z_2$ toric code, which is an Abelian topological order.

\begin{table*}[htb]
\begin{center}
\includegraphics[width= 1.9\columnwidth]{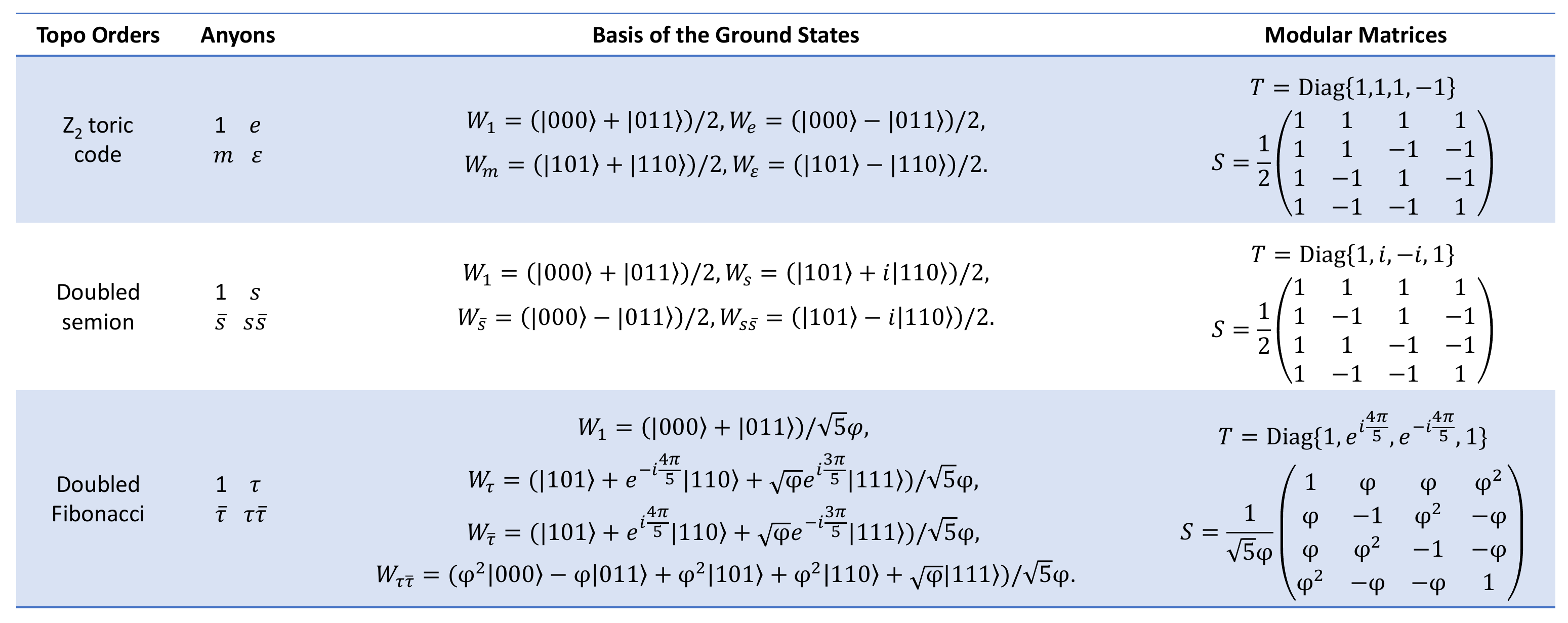}
\end{center}
\setlength{\abovecaptionskip}{-0.00cm}
\caption{\footnotesize{Anyon types, the basis of the ground states, and modular matrices $T$ and  $S$ for the three topological orders: $\Z_2$ toric code, doubled semion, and doubled Fibonacci, of the minimal honeycomb lattice on a torus. With respect to the $T$ and  $S$  matrices, the rows and columns are in the order of the anyon types listed in the second column. For the non-Abelian doubled Fibonacci order, the parameter $\varphi=(1+\sqrt{5})/2$, the golden ratio.}}\label{topoorder}
\end{table*}

2. \textit{Doubled semion}. The input data set of this topological order differs from that of the $\Z_2$ toric code by $d_1=-1$.
The four types of anyons are $1$, $s$, $\bar s$, and $s\bar s$, among which $s$ and $\bar s$ are semions.
A semion may be thought as a half fermion because its statistics is $i$ instead of $-1$. The set of output data characterizes the doubled semion order, which is also an Abelian topological order.

\begin{figure}[!ht]
\centering
\subfigure[]{\label{fig:honeycomb}
\includegraphics[scale=1.5]{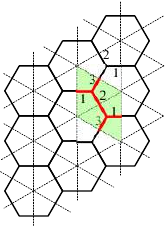}
}
\subfigure{\raisebox{45pt}{$\xRightarrow[\text{counterclockwise}]{\frac{\pi}{3}\ \text{rotation}}$}
}
\subfigure[]{\label{fig:honeycombR}\addtocounter{subfigure}{-1}
\includegraphics[scale=1.5]{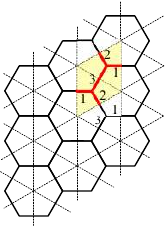}
}
\caption{(a) Honeycomb lattice with periodic boundary condition on the unit cell (green region) consisting of three edges $1,2$, and $3$. This is in fact the minimal honeycomb lattice on a torus in Fig. \ref{fig:minTorus}. (b) Unit cell (yellow region) obtained from (a) by a $\pi/3$ rotation counterclockwise.}
\label{fig:honeycombAll}
\end{figure}

3. \textit{Doubled Fibonacci}.
We still have two string types $0$ and $1$, with, however, a new fusion rule $1\x 1=0\x 1$. This fusion rule leads to a different set of $F$-symbols, such that $d_0=1$ and $d_1=\varphi=(1+\sqrt{5})/2$, the golden ratio.
The four types of anyons are $1$, $\tau$, $\bar \tau$, and $\tau\bar\tau$, respectively. The anyon $\tau$ is called the Fibonacci anyon because the dimension of the Hilbert space of $n$ $\tau$'s grows as the Fibonacci sequence with $n$\cite{Nayak2008}. The anyon $\bar\tau$ is the same as $\tau$ except that it has an opposite self-statistics. So, the anyon $\tau\bar\tau$ is a bound state of $\tau$ and $\bar\tau$. This is why the output topological order is called the doubled Fibonacci. The fusion of two Fibonacci anyons is $\tau\x\tau=1+\tau$; hence, the Hilbert space of two Fibonacci anyons is two-dimensional and can be identified as the space of a logical qubit. The more Fibonacci anyons excited, the larger the logical space. The Fibonacci anyons are non-Abelian, whose braiding fabricates unitary quantum gates well suited for universal TQC \cite{Stern2006,Nayak2008}.

\textit{Experimental implementation ---}
Our goal is to simulate the above three topological orders by 1) preparing their ground states, 2) performing the modular transformation on the virtual minimal honeycomb lattice, and 3) measuring the modular matrices that can uniquely distinguish the three topological orders. Here, we show how we would perform the modular transformations. It turns out that on the honeycomb lattice on a torus, $T$ and  $S$ matrices cannot be simultaneously measured; however, it is shown that a $\pi/3$ rotation of the lattice about the axis perpendicular to the lattice surface is equivalent to performing the combined modular transformation $ST^{-1}$ \cite{Cincio2013}. What crucial is that the three topological orders possess distinct matrices $ST^{-1}$ and thus can be distinguished by measuring these matrices. Hence, we just need to know how a $\pi/3$ rotation acts on the ground states of a topological order. On a torus, a $\pi/3$ rotation transforms the minimal honeycomb lattice as depicted in Fig. \ref{fig:honeycombAll}. It is easy to see that the rotation cyclically permutes the three edges by $1\rightarrow 2\rightarrow3\rightarrow 1$. Consequently, the rotation transforms any state by $\ket{e_1 e_2 e_3}\rightarrow \ket{e_3 e_1 e_2}$, implying that one can then apply this permutation operation to the three bases of ground states and generate three new bases. Having done this, it is then straightforward to show that the inner product between the new and original bases reproduces the modular matrices $ST^{-1}$. In our experiment, the effect of the $\pi/3$ rotation, i.e., the cyclic permutation $1\rightarrow 2\rightarrow3\rightarrow 1$, is implemented by two SWAP gates, SWAP$_{12}$ and SWAP$_{23}$.

\begin{figure}[htpp]
    \centering\includegraphics[width=0.96\columnwidth]{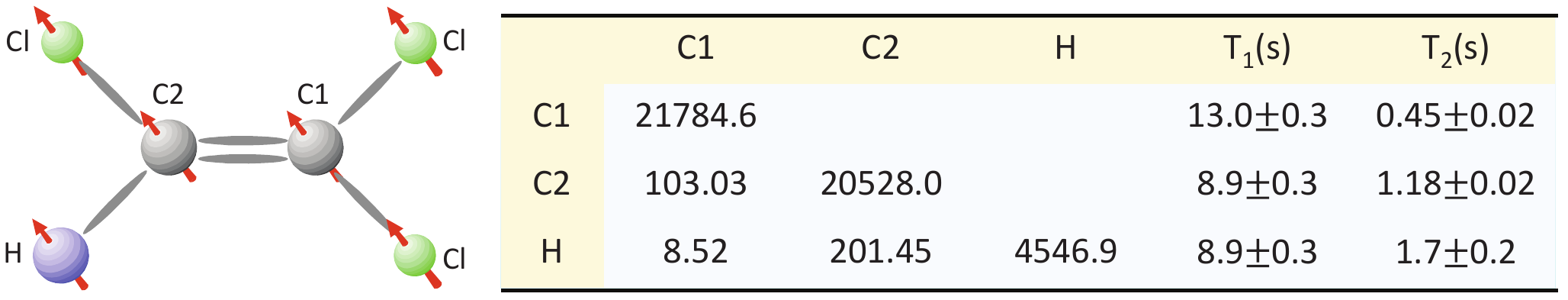}
    \caption{Molecule structure of TCE, where one $^{1}$H and two $^{13}$C's form a three-qubit system. The table on the right lists the parameters of chemical shifts (diagonal, Hz), J-coupling strengths (off-diagonal, Hz), and relaxation timescales T$_1$ and T$_2$ (second). }\label{tce}
\end{figure}

Our $3$-qubit system is represented by the $^{13}$C-labeled  trichloroethylene (TCE) molecule dissolved in d-chloroform \cite{lu2014experimental}. The sample consists of two $^{13}$C's and one $^{1}$H, as shown in Fig. \ref{tce}. All parameters of the molecule are listed in the table of Fig. \ref{tce}, and all experiments are carried out on a Bruker DRX 700MHz spectrometer at room temperature.

Each experiment of simulating a given topological order was divided into the following three steps. Certain details can be found in the supplemental material \cite{supple}.

1) Prepare the ground states. We first created a pseudo-pure state (PPS) \cite{cory1997ensemble} with the experimental fidelity over 0.99, and then prepared it into one of the ground states for $\Z_{2}$ toric code, doubled semion, and doubled Fibonacci order as shown in Table. \ref{topoorder}, respectively. Ground-state preparation is realized by the gradient ascent pulse engineering (GRAPE) optimizations \cite{khaneja2005optimal,ryan2008liquid}, with each pulse 10 ms. Denote each ground state of the currently simulated topological order as $\ket{\phi_{i}}$ ($1\leq i\leq 4$).

2) Perform the modular transformation. For each of the four ground states $\ket{\phi_{i}}$, we apply two SWAP gates between qubit 1 and 2, and then qubit 2 and 3, to cyclically permute the three qubits. They were optimized by the GRAPE technique with pulse durations of 20 ms. It is equivalent to performing the modular transformation ($\pi/3$ rotation) on the torus of the minimal honeycomb lattice. Denote each new ground state of the currently simulated topological order as $\ket{\psi_{i}}$ ($1\leq i\leq 4$).

3) Measure the ground states before and after the modular transformation. To acquire the $ST^{-1}$ matrix in experiment, we need to calculate the inner products between the original and new ground states. A full state tomography was implemented before and after the modular transformations, to obtain the information of the original ground states $\ket{\phi_{i}}$ and new ground states $\ket{\psi_{i}}$, respectively.

\begin{figure}[htpp]
    \centering\includegraphics[width=0.9\columnwidth]{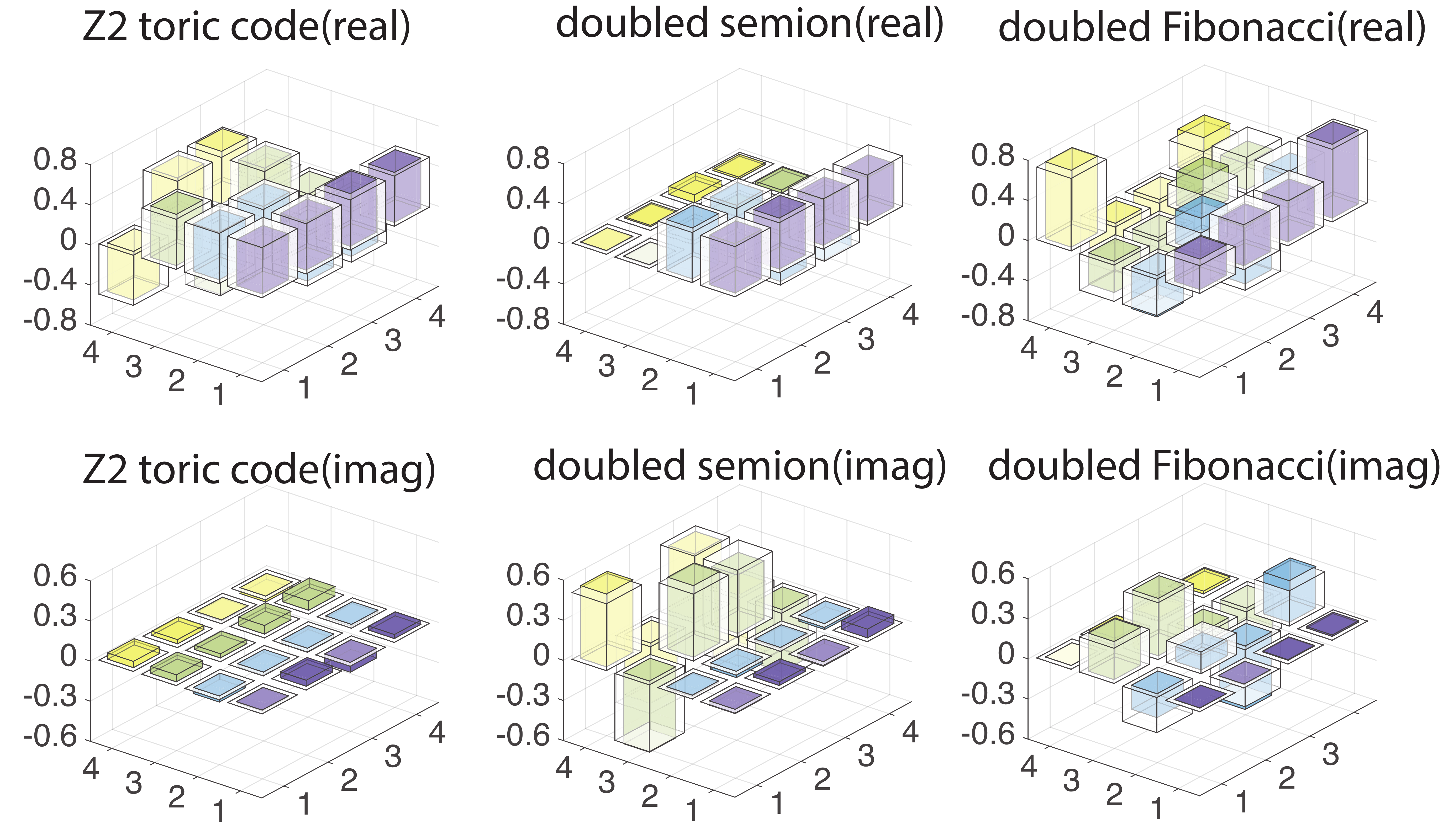}
    \caption{$ST^{-1}$ matrices for $\Z_{2}$ toric code, doubled semion, and doubled Fibonacci topological orders, respectively.  The transparent columns represent the theoretical values, and the colored represent the experimental results.}\label{stmat}
\end{figure}

Note that the state tomography inevitably leads to mixed states in experiment for the sake of experimental errors. To calculate the inner products of the two ground states, it is necessary to purify the measured density matrices to pure states. This purification step was realized by the maximum likelihood method \cite{vrehavcek2001iterative}, and say $\ket{\phi_{i}^{\text{exp}}}$ and $\ket{\psi_{i}^{\text{exp}}}$ were found to be the closest to our experimental density matrices. As a result, each element in the experimentally reconstructed $ST^{-1}$ matrix was
\begin{equation}
ST^{-1}_{ij}=\langle\phi_{i}^{\text{exp}} |\psi_{j}^{\text{exp}} \rangle,
\end{equation}
from which the entire $ST^{-1}$ could be reconstructed.

In Fig. \ref{stmat}, all the $ST^{-1}$ matrices of the $\Z_{2}$ toric code, doubled semion and doubled Fibonacci topological orders are illustrated. The real parts of $ST^{-1}$ are displayed in the upper row, and the imaginary in the lower row. In each figure, the transparent columns stand for the theoretical values, and the colored stand for the experimental results. From the figure, we conclude that our experiment matches well with the theoretical predictions, and each topological order is indeed identified clearly from its measured $ST^{-1}$ matrix.

We also calculated the average fidelity \cite{emerson2005scalable} between the theoretical $ST^{-1}$ matrix and the experimental one. For the non-Abelian doubled Fibonacci topological order, the average fidelity is $0.983\pm 0.005$, while for the other two Abelian topological orders $\Z_{2}$ toric code and doubled semion, the average fidelities are $0.993\pm0.002$ and $0.992\pm0.003$, respectively. This provides another evidence that we have successfully identified distinct topological orders with high confidence using our quantum simulator.

We clarify that state tomography is not necessary in measuring the modular matrices of topological orders, if an ancilla qubit is involved and the modular transformations are modified correspondingly \cite{miquel2002interpretation}. To guarantee the experimental precision, in this work we used a three-qubit simulator and implemented full state tomography. For how to measure the modular matrices in a tomography-free way, see the supplemental material \cite{supple}.

\textit{Discussion} --- TQC is undoubtedly a very promising scheme of quantum computing, which requires the engineering of Hamiltonians with many-body interactions. Due to the notorious difficulties in engineering such Hamiltonians experimentally, most of the preliminary experiments towards TQC adopted a state preparation approach \cite{han2007scheme} to demonstrate the exotic properties of anyons, such as the fractional statistics \cite{lu2009demonstrating,pachos2009revealing,feng2013experimental,zhong2016emulating} or path independence \cite{park2016simulation}. Each experimental platform has its own advantages and drawbacks. For example, the photonic system \cite{lu2009demonstrating,pachos2009revealing} has genuine entanglement, but generates the states probabilistically which is inefficient; the NMR system \cite{feng2013experimental,park2016simulation} has good controllability, but it is lack of entanglement and scalability; the superconducting circuit \cite{zhong2016emulating} is a solid-state system with genuine entanglement, but it requires extremely low temperature. In addition, the state preparation approach \cite{han2007scheme} just provides a way to mimic anyonic properties, which is not suited to verify the robustness of TQC. Recently, four-body ring-exchange interactions and anyonic excitations were observed in ultracold atoms \cite{dai2016observation}, which is an essential step towards the realization of TQC.

The experiments mentioned above account for the toric code only, which is Abelian. Universal TQC nonetheless requires non-Abelian anyons, for the Fibonacci order as the simplest example. Our work is by far the first experimental measurement of the modular matrices of topological orders, in particular non-Abelian topological orders, and thus opens up a way to measure modular matrices using quantum simulators.

\textit{Conclusion} --- Echoing the equivalence between the quantum circuit scheme and the topological quantum computation scheme, on a NMR quantum simulator, we successfully identify the doubled Fibonacci topological order, which is a promising candidate for topological quantum computation. Since the doubled Fibonacci order is one of the three topological orders described by the string-net model with two string types, using the same system, we also identify the other two topological orders, i.e., the $\Z_2$ toric code and the doubled semion. Our simulator can serve as a basic module for simulating the dynamical properties --- in particular braiding and edge effects --- of these topological orders.

\begin{acknowledgements}
We are grateful to the following funding sources:   NSERC, Industry Canada, and CIFAR (D.L., B.Z., and R.L.); National Natural Science Foundation of China under Grants No. 11175094 and No. 91221205 (K.L., and G.L.); National Basic Research Program of China under Grant No. 2015CB921002 (K.L., and G.L.). Y.W. thanks Lukasz Cincio, Yuting Hu, Chenjie Wang and Yang Qi for helpful discussions. Y.W. is supported by the Fudan University startup grant. He was also supported by the John Templeton foundation No. 39901, where part of this work was done. This research was supported in part by Perimeter Institute for Theoretical Physics. Research at Perimeter Institute is supported by the Government of Canada through the Department of Innovation, Science and Economic Development Canada and by the Province of Ontario through the Ministry of Research, Innovation and Science. K. L., Y. W., and L. H. contributed equally to this work.
\end{acknowledgements}

\end{document}